\documentclass{emulateapj}
\usepackage{apjfonts}\tighten

\newcommand\Iacs{\ifmmode I_{814}\else$I_{814}$\fi}
\newcommand\racs{\ifmmode r_{625}\else$r_{625}$\fi}
\newcommand\gacs{\ifmmode g_{475}\else$g_{475}$\fi}
\newcommand\kms{km~s$^{-1}$}

\newcommand\etal{{et~al.}} 
\newcommand\mM{\ifmmode(m{-}M)\else$(m{-}M)$\fi}
\newcommand\msun{\ifmmode{M_\odot}\else{$M_\odot$}\fi}
\newcommand\hst{{\it HST}}
\newcommand\rIcolor{{\ifmmode{(r_{625}{-}I_{814})}\else$(r_{625}{-}I_{814})$\fi}}
\newcommand\gIcolor{{\ifmmode{(g_{475}{-}I_{814})}\else$(g_{475}{-}I_{814})$\fi}}
\newcommand\lta{\lesssim}
\newcommand\gta{\gtrsim}
\newcommand\taul{\ifmmode\tau_{L}\else$\tau_{L}$\fi}
\newcommand\rc{\ifmmode R_{e,{\rm c}}\else$R_{e,{\rm c}}$\fi}
\def\magauto{{\sc mag\_auto}}
\def\txitxo{Ben\'{\i}tez}
\def\sersic{S\'{e}rsic}
\def\esog{ESO\,325-G004}
\def\hasegan{Ha{\c s}egan}
\def\cote{C\^ot\'e}
\shortauthors{Blakeslee \& Barber DeGraaff}
\shorttitle{UCDs in Abell S0740}
\slugcomment{Accepted for Publication in AJ}

\begin{document}

\title{Ultra-Compact Dwarf Candidates Near the Lensing Galaxy 
in Abell S0740\altaffilmark{1}}

\author{John P.~Blakeslee\altaffilmark{2,3} and Regina Barber DeGraaff\altaffilmark{2}}
\altaffiltext{1}{Based on observations made with the NASA/ESA Hubble
Space Telescope, obtained from the Space Telescope Science Institute,
which is operated by the Association of Universities for Research in
Astronomy, Inc., under NASA contract NAS\,5-26555.
These observations are associated with program \#10429.}
\altaffiltext{2}{Department of Physics \& Astronomy, Washington State
  University, Pullman, WA 99164}
\altaffiltext{3}{Herzberg Institute of Astrophysics, 5071 West Saanich Road,
Victoria, BC V9E\,2E7, Canada; john.blakeslee@nrc.ca}

\begin{abstract}
We analyze three-band imaging data of the giant elliptical galaxy \esog\ from
the \textit{Hubble Space Telescope} Advanced Camera for Surveys (ACS).  This
is the nearest known strongly lensing galaxy, and it resides in the center of
the poor cluster Abell S0740 at redshift $z{\,=\,}0.034$.  Based on magnitude,
color, and size selection criteria, we identify a sample of 15 ultra-compact
dwarf (UCD) galaxy candidates within the ACS field.  This is comparable to the
numbers of UCDs found within similar regions in more nearby clusters (Virgo,
Fornax, Hydra).  We estimate circular half-light radii $\rc$ from 2-D
\sersic\ and King model fits and apply an upper cutoff of 100~pc for our UCD
selection.  The selected galaxies have typical \sersic\ indices $n{\,\approx\,}1.5$,
while larger sources with $\rc{\;>\,}100$~pc are more nearly exponential, 
perhaps indicating that the latter are dominated by background disk galaxies. 
Many of the UCD candidates are surrounded by a faint ``fuzz'' of
halo light, which may be the remnants of stripped material, and there is some
evidence for intrinsic flattening of the UCDs themselves.  An apparent
separation in size between the most compact UCDs with $\rc<17$~pc and larger
ones with $\rc>40$~pc may hint at different formation mechanisms.
We do not find any M32 analogues in this field.  
The colors of the UCD candidates span the range from blue to red globular
clusters, although the brightest ones are predominantly red.
The UCD candidates follow the flattened, elliptical distribution of the
globular clusters, which in turn follow the galaxy halo light, suggesting a
common evolution for these three components.
Planned follow-up spectroscopy can determine which candidates are truly members 
of Abell S0740 and how similar they are in distribution to the globulars.
\end{abstract}
\keywords{galaxies: elliptical and lenticular, cD ---  
galaxies: dwarf ---  
galaxies: evolution ---
galaxies: individual (ESO 325-G004)  ---
galaxies: clusters: individual (Abell~S0740)
\vspace{0.2cm}
}

\section{Introduction}
A new class of stellar system has emerged in recent years.  Due to the size
of these objects, being larger than average globular clusters (GCs) and smaller
than dwarf galaxies, they have been dubbed ultra-compact dwarf galaxies,
or UCDs (Phillips \etal\ 2001).
They are typically a few $\times 10^7$ \msun\ in mass, with effective radii in
the range 10-100~pc.
First discovered in the Fornax Cluster (Hilker et al.\ 1999; Drinkwater et
al.\ 2000), UCDs have now been found in significant numbers in the Virgo,
Centaurus, and Hydra clusters (\hasegan\ \etal\ 2005; Mieske et al 2007;
Wehner \& Harris 2007), all systems within $\sim\,$50~Mpc of the Local Group.
They are apparently very rare outside of galaxy clusters 
(Evstigneeva \etal~2007b).

As they have absorption line spectra and appear to be transitional between
GCs and early-type dwarfs (cf. \hasegan\ \etal\ 2005), there are two basic
ideas for the nature of UCDs: they are related to globulars, or to dwarf
galaxies.  More specifically, UCDs may be the largest members of the rich GC
populations found inside galaxy clusters (Mieske \etal\ 2002), possibly
growing to such large size through dissipational merging early in their
lifetimes (Fellhauer \& Kroupa 2002).  Or, they may be the small, tidally
stripped remains of nucleated dwarf galaxies on orbits that carried them too
close to the center of the cluster potential (Bekki \etal\ 2001; Drinkwater
\etal\ 2003).  This latter explanation has come to be known as galaxy
``threshing,'' but the idea has been around for many years.  
Bassino \etal\ (1994) numerically simulated the evolution of nucleated dwarfs
in Virgo and showed that stripped nuclei could constitute a large
fraction of M87's very rich GC system, while larger UCD-like remnants would
occur farther out.  Recent \textit{Hubble Space Telescope} (\hst)
imaging has revealed nuclei in a much higher percentage of Virgo
early-type dwarfs than previously thought (\cote\ \etal\ 2006).  Thus,
stripping of nucleated dwarfs may account for both UCDs and many of the
GCs in the center of cluster potentials.

\begin{figure*}
\epsscale{0.92}
\plotone{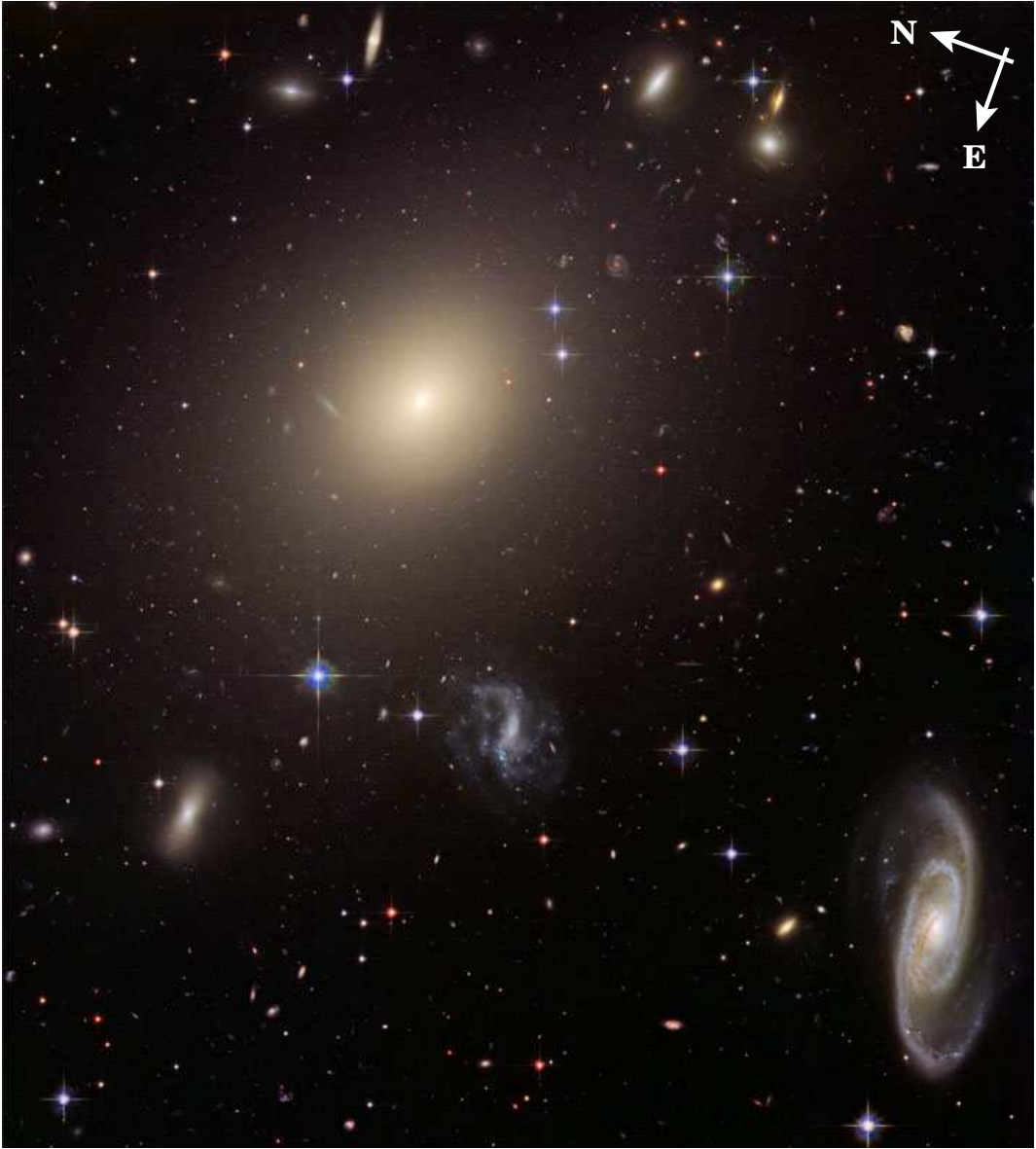}
\caption{\textit{Hubble Space Telescope} ACS/WFC image 
  of ESO\,325-G004, showing about $3\farcm0{\,\times\,}3\farcm3$ of the
  field at the observed orientation.  This color composite was
  constructed by the Hubble Heritage Team (STScI/AURA)
  from our imaging in the F475W $(g)$,
  F625W $(r)$, and F814W $(I)$ bandpasses.}
\label{fig:pic}
\end{figure*}

Some evidence based on the color-magnitude sequence of UCDs suggests that
they may be an extension of the red GC component to brighter magnitudes
(Wehner \& Harris 2007).  The UCDs in the Virgo and Fornax clusters also have
spectroscopic metallicities and $\alpha$-element enhancements consistent with
their being the high-mass mass extreme of the red GC population (Evstigneeva
\etal\ 2007a; Mieske et al.\ 2006), and less consistent with simple versions
of the threshing model.
Estigneeva \etal\ (2008) attempted to distinguish between the two formation
scenarios on the basis of the structural properties of UCDs in the nearby
Virgo and Fornax clusters measured using the \hst\ High Resolution Channel.
Even with such high resolution measurements, the data were consistent with
either explanation, although more detailed predictions of the size evolution
of the nuclei during threshing are needed to test this scenario.  The
relatively low velocity dispersions of cluster UCD populations are expected in
either model (e.g., Bekki 2007).  However, detailed comparison between the
spatial distributions of large samples of UCDs and their possible nucleated
dwarf progenitors in clusters may help
uncover their evolutionary histories (e.g., Goerdt \etal\ 2008; Thomas
\etal\ 2008).

Given the difficulty in distinguishing between the formation scenarios,
further UCD surveys can provide valuable information on the properties of this
new type of stellar system.  A larger sample of groups and clusters is
especially useful for constraining environmental effects on the formation of
UCDs.  Here, we present a search with \hst\ for UCD candidates near \esog,
the central giant elliptical in Abell S0740.  This is one of
the systems in the supplementary list of  poor clusters 
tabulated by Abell \etal\ (1989) that did not meet the lowest richness 
criteria of the original Abell (1958) catalogue.  
The cluster velocity dispersion is only $\sim\,$300 \kms\ (see plot in Smith
\etal\ 2005), similar to that of Fornax, where UCDs were first discovered.
The absolute $V$ magnitude of \esog\ is $M_V=-23.2$, making it about 60\%~more
luminous than M87, or 2.5 times the luminosity of NGC\,1399 in Fornax.  At
$z=0.034$, \esog\ is also the closest known gravitational lens and has both
dynamical and lensing mass estimates (Smith \etal\ 2005). This makes it an
interesting target for UCD searches, since it is a very massive, dominant
elliptical in a poor cluster or rich group environment.

The following section describes our data in detail.  In
\S\,\ref{sec:selection}, we present our photometric and size measurements and
discuss the selection of UCD candidates.  The properties of the UCD candidates
are discussed in \S\,\ref{sec:props} and compared with those of GCs and other
objects in the field.  The final section summarizes the results.
Throughout this paper, we use the WMAP 3-year cosmology results (Spergel et
al. 2007) and assume a distance modulus for \esog\ of $(m{-}M) = 35.78$ mag, or a
luminosity distance of 143 Mpc, and an angular scale of 0.65 kpc
arcsec$^{-1}$.  This translates to an image scale of about 33 pc pix$^{-1}$
for our observations with Advanced Camera for Surveys Wide Field Channel
(ACS/WFC).

\section{Observations and Reductions}
\label{sec:data}

\esog\ was imaged with the ACS/WFC in the F475W, F625W, and F814W filters.
Throughout this paper, we refer to magnitudes in these filters as 
\gacs, \racs, and \Iacs, respectively.
The galaxy was initially observed in F814W and F475W as part of 
\hst\ GO Program 10429 during January 2005.
This program, which is conducting a surface brightness fluctuation 
survey in the Shapley supercluster region, is described in Blakeslee (2007).
There were 22 F814W exposures of varying times totaling of 18,882~s, and
three exposures in F475W of 367s each.
In February 2006, further imaging of the \esog\ field 
was carried out by \hst\ DD Program 10710 for a Hubble Heritage 
public release image.\footnote{http://heritage.stsci.edu/2007/08/index.html}
This provided six additional exposures in each of the F475W and F625W filters.
The total exposure times for this field were therefore 
5901, 4650, and 18882~s in F475W, F625W, and F814W, respectively.

The images were processed with the Apsis pipeline (Blakeslee \etal\
2003) to produce summed, geometrically corrected, cosmic ray cleaned
images for each bandpass.  Figure~\ref{fig:pic} shows a color composite image
constructed from the data in the three bandpasses.  Apsis corrects the astrometric zero
point of the images to within an uncertainty of about 0\farcs1.
It also produces an RMS image giving the total noise for each pixel.
We calibrated the photometry using the Vega-based ACS/WFC
zero points for each filter from Sirianni \etal\ (2005): 
$m_{\gacs}=26.168$, $m_{\racs}=25.731$, and $m_{\Iacs}= 25.501$.  
We corrected the photometry for Galactic extinction using $E(B-V)= 0.0605$
mag (Schlegel \etal\ 1998) and the extinction ratios from Sirianni \etal\
(2005).  We find the following extinction corrections in each band: 
$A_{475} = 0.217$ mag, and $A_{625} = 0.159$ mag, and $A_{814} = 0.109$ mag.


We modeled the main galaxy \esog\ using the elliprof software (Tonry \etal\
1997), as well as several of the other smaller galaxies in the field to obtain
a better fit.  The small galaxy models were subtracted from the image, and
bright stars, diffraction spikes, and other galaxies were masked 
so a final model of \esog\ could be made.  This final model was then
subtracted to create a residual image, which was used to find sources with the
object detection software SExtractor (Bertin \& Arnouts 1996).  We used the
Apsis RMS image, which includes the noise from the subtracted galaxies, for
the SExtractor detection weighting.  To the F814W RMS image, we also added
additional noise to account for the galaxy surface brightness
fluctuations, as described in more detail by Jord\'an \etal\ (2004) and
Barber DeGraaff \etal\ (2007).   We used SExtractor in ``dual image mode'' with
the much deeper F814W image as the detection image in each case, and
individual filter images used for the photometry.  This ensures that the same
object centroids and measurement apertures are used for all the images,
resulting in the most accurate color measurements (see \txitxo\ \etal\ 2004).
We adopt the SExtractor \magauto\ value for the total \Iacs\ magnitude and
isophotal magnitudes to measure galaxy colors.

\begin{figure*}
\epsscale{0.9}
\plottwo{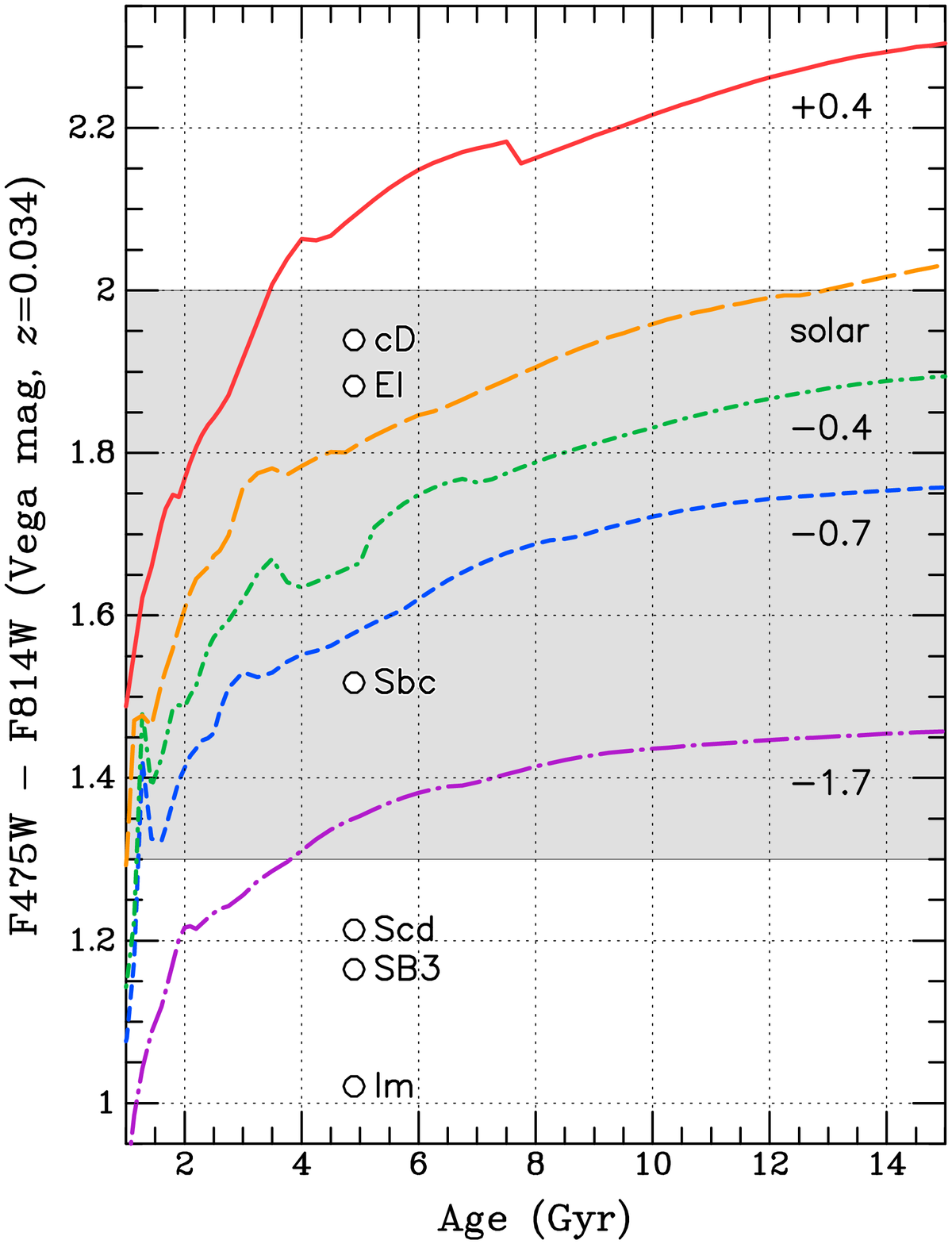}{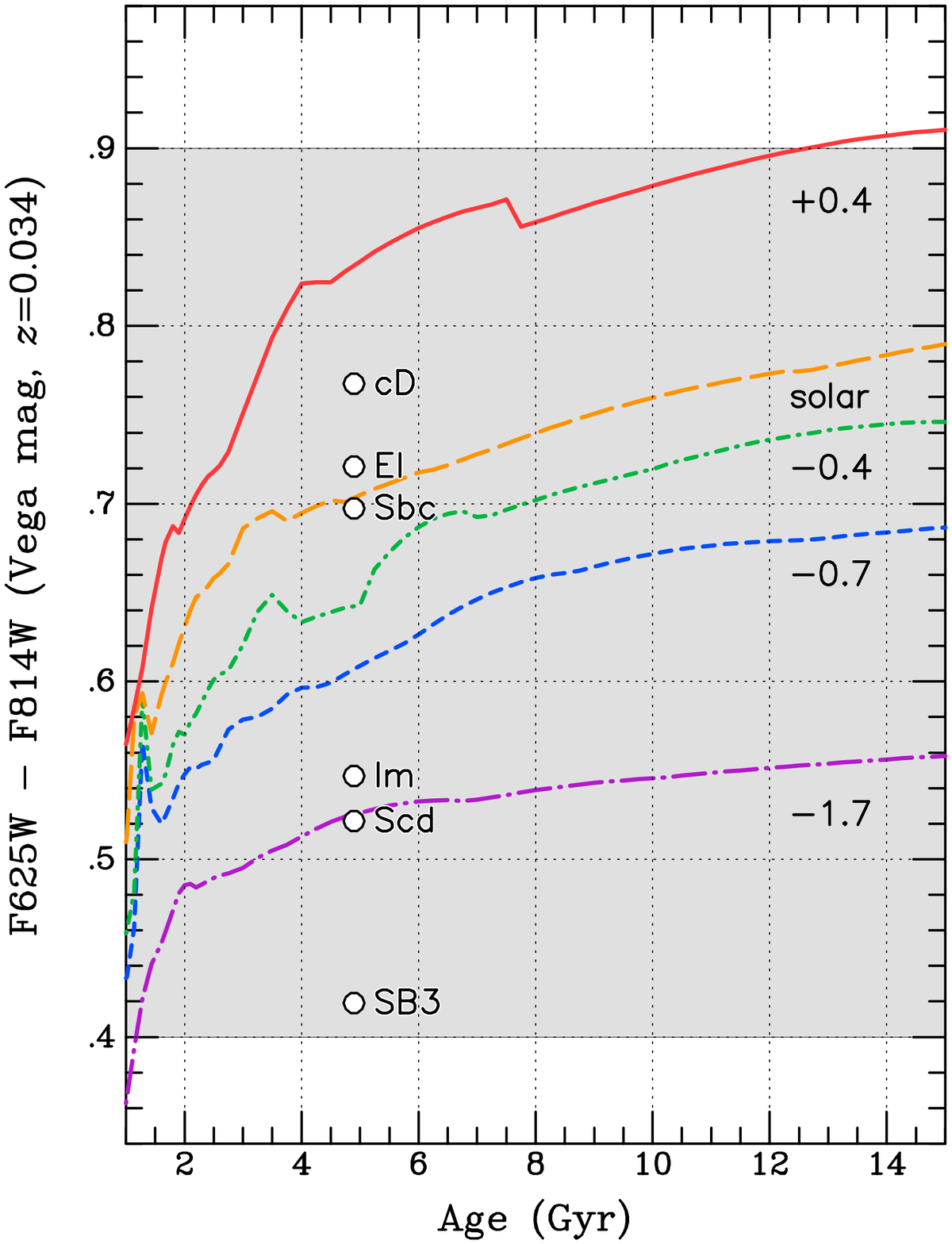}
\caption{%
Predicted age evolution in the observed ACS colors at redshift
$z{=}0.034$ for Bruzual \& Charlot (2003) single-burst stellar population
models with five different metallicities, labeled by their [Fe/H] values.
We also show the expected colors at this redshift for six different
empirical galaxy templates (see text) with arbitrary placement along the
horizontal axis.
The shaded areas delineate the color selection criteria for the UCD candidates.
The broader baseline \gIcolor\ color is used for the more stringent selection
cut, based on the expected range of stellar populations in UCDs. 
The less-sensitive \rIcolor\ cut is simply to ensure the objects have
reasonable colors for galaxies at this redshift.}
\label{fig:colorage}
\end{figure*}

\section{Sample Selection}
\label{sec:selection}

\subsection{Color and Magnitude Cuts}
\label{sec:cmag}

In order to search for UCDs in this field, we first applied cuts in color and
magnitude to select a list of objects for size and shape measurements.  Large
numbers of GCs are visible in the image, but we expect the turnover, or peak
luminosity, of the GC luminosity function (GCLF) to occur near
$\Iacs\approx27.5$.  Therefore, only $\sim\,$1\% of the GCs should have magnitudes
$\Iacs<24$, which is about 2.7\,$\sigma$ brighter than GCLF peak, and we
chose $\Iacs=24$ as the faint limit for UCD candidates. This corresponds to
an absolute $V$ magnitude $M_V\approx-10.8$, which typically marks the
transition between GCs and UCDs (e.g., \hasegan\ \etal\ 2005).  However, we performed
the surface photometry fits and size measurements described below to a limit 
one magnitude fainter than this.

To derive color cuts, we calculated the color evolution for Bruzual \& Charlot
(2003) simple stellar population (SSP) models in the observed bandpasses at
$z=0.034$, as well as the colors of empirical galaxy templates from \txitxo\
\etal\ (2004) and NGC\,4889 in the Coma cluster, which we use as a
template cD galaxy.
Figure~\ref{fig:colorage} shows the results of these calculations.  The
broader baseline \gIcolor\ has more discriminating power, and we use it for
our more stringent color selection criterion: $1.3 < \gIcolor < 2.0$, which
corresponds to $0.85 < V{-}I < 1.35$, based on the models.  This range 
includes the photometrically transformed colors of \textit{all}
confirmed UCDs from previous studies (e.g., Mieske \etal\ 2004b, 2007;
\hasegan\ \etal\ 2005, Evstigneeva \etal\ 2008).  The color cut spans the
range from Sc-type spirals to the reddest giant ellipticals, and from
intermediate age, very metal-poor models to metal-rich models.  Note that the
models do not include alpha-enhancement, and the absolute metallicity scale
should be viewed as approximate; the empirical templates are the more useful
comparison.

Additionally, we require $0.4 < \rIcolor < 0.9$, a broad cut which simply
ensures that the objects have reasonable colors for a galaxy at this redshift.
We also attempted to use our multi-band imaging data to estimate
photometric redshifts as part of the selection criteria, similar to Mieske
\etal\ (2004a) who searched for UCD-like objects in the more distant cluster
Abell~1689 and had the benefit of a fourth bandpass.  However, we found that
the photometric redshifts based on just these three bands were not very robust
for objects in this low-redshift cluster.  We therefore decided to use the
simple color cuts highlighted in Figure~\ref{fig:colorage}.  No additional
objects would be included in our final sample of best UCD candidates if we
relaxed the \gIcolor\ color cut to a very red limit of~2.2.

\begin{figure*}
\epsscale{1.05}
\plottwo{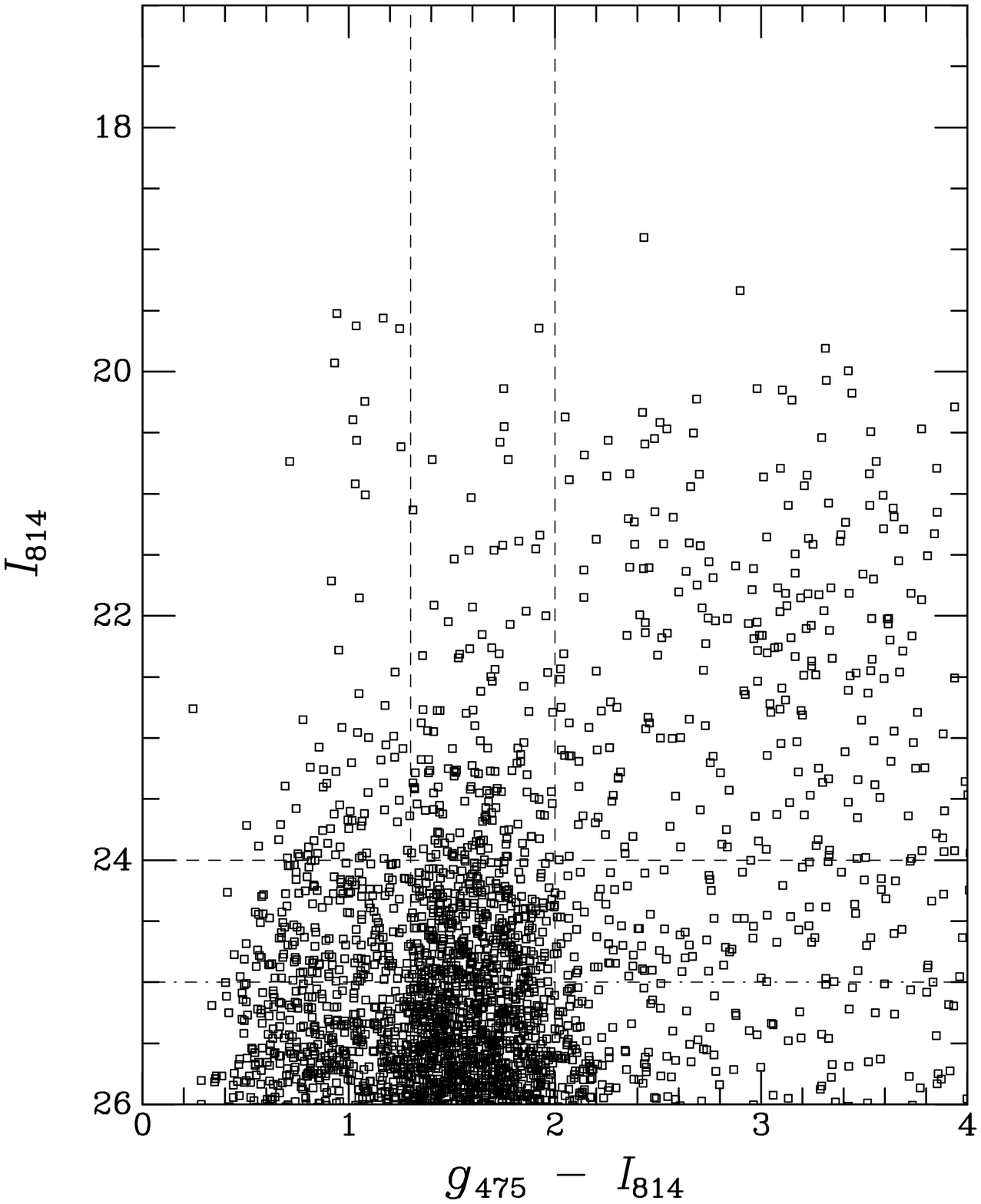}{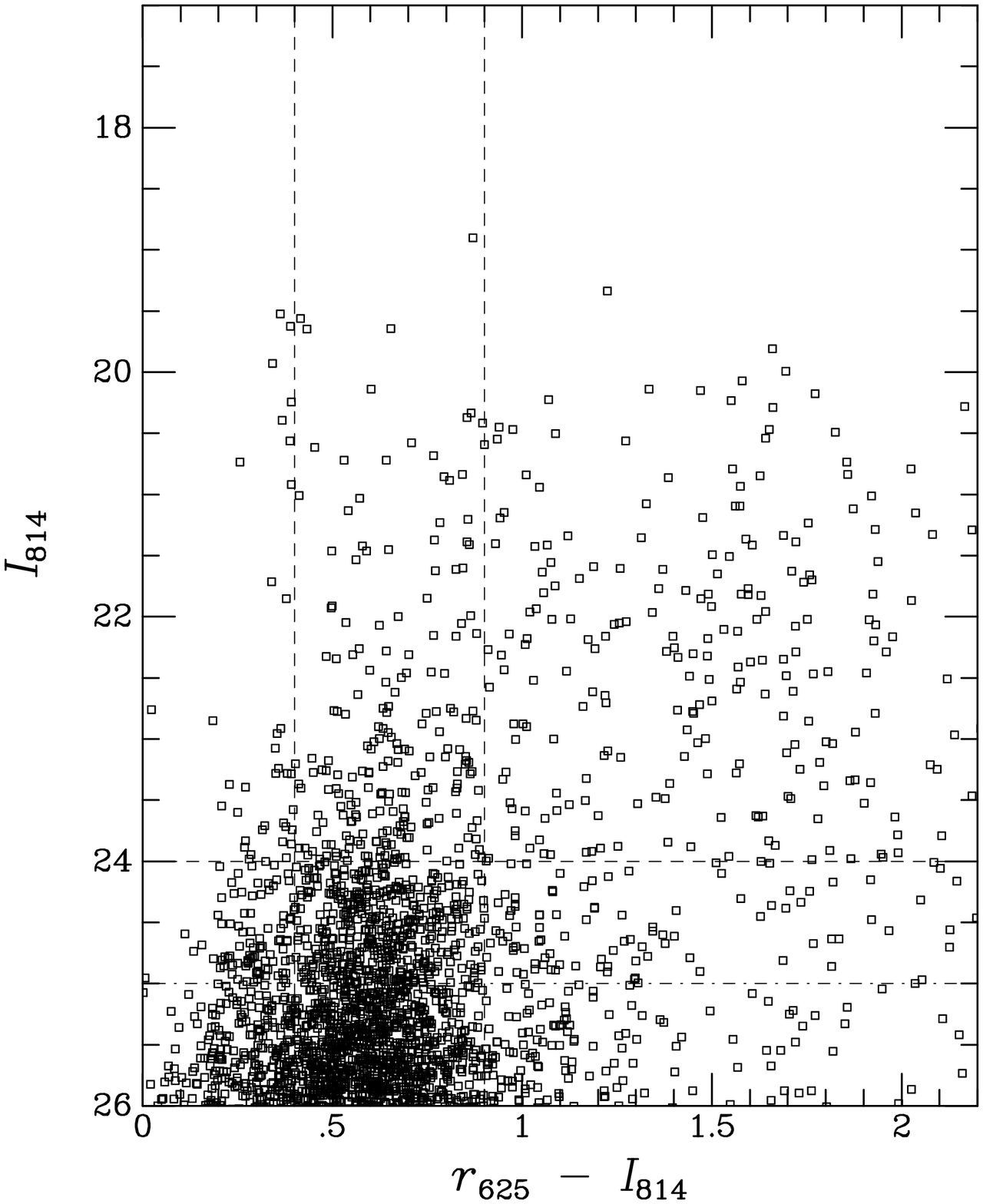}
\caption{Color-magnitude diagrams for all objects detected in our images down to
$\Iacs = 26$.  The vertical dashed lines show the color cuts from 
Figure~\ref{fig:colorage}.  The horizontal dashed line at $\Iacs=24$ shows the
faint limit we impose for UCD candidates; fainter than this, the objects at
these colors are mainly globular clusters in \esog.  
The dot-dashed horizontal line at $\Iacs=25$
is the limit we use for the 2-D surface photometry fits.
}\label{fig:all}
\end{figure*}

\begin{figure*}
\epsscale{1.05}
\plottwo{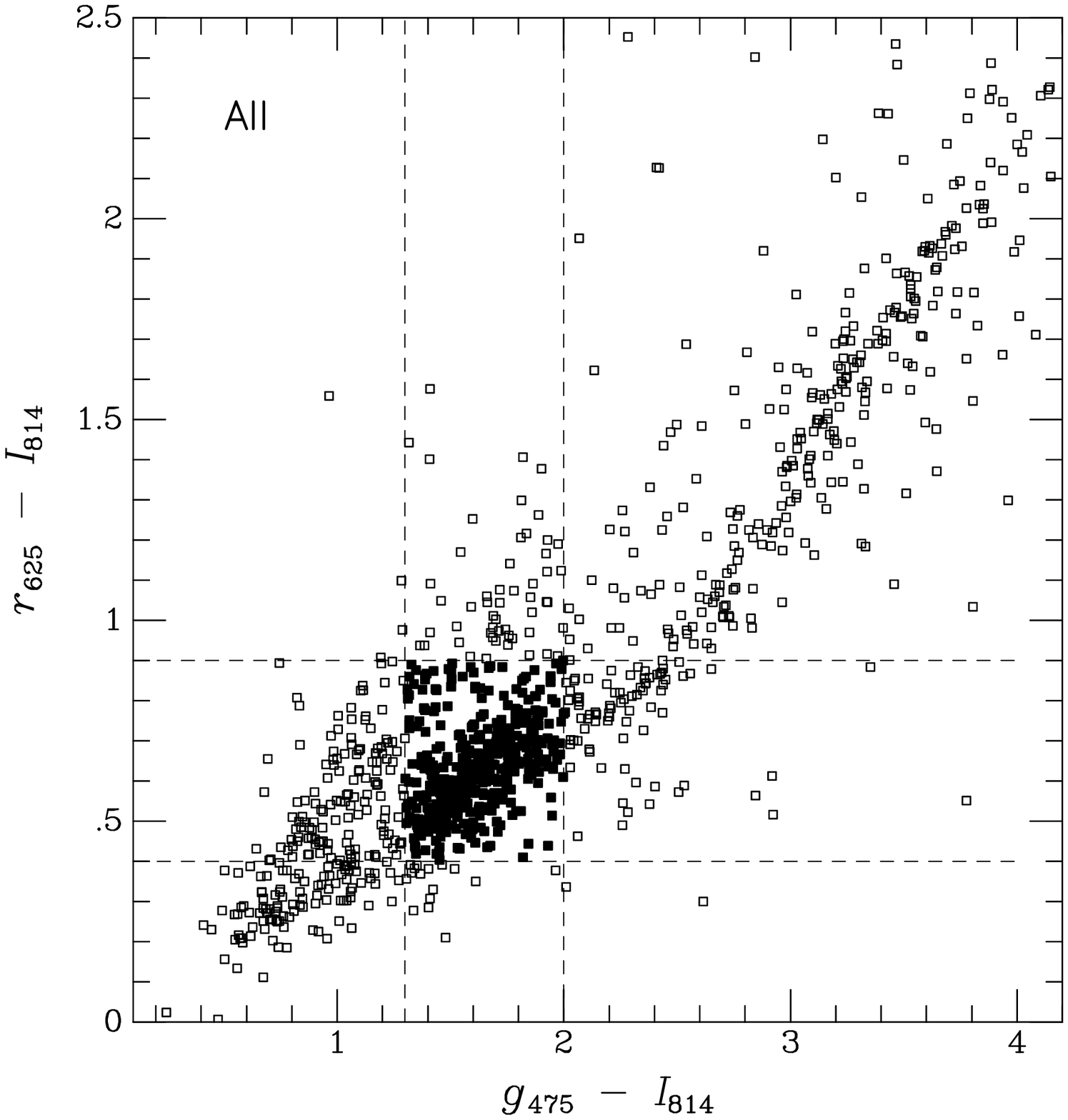}{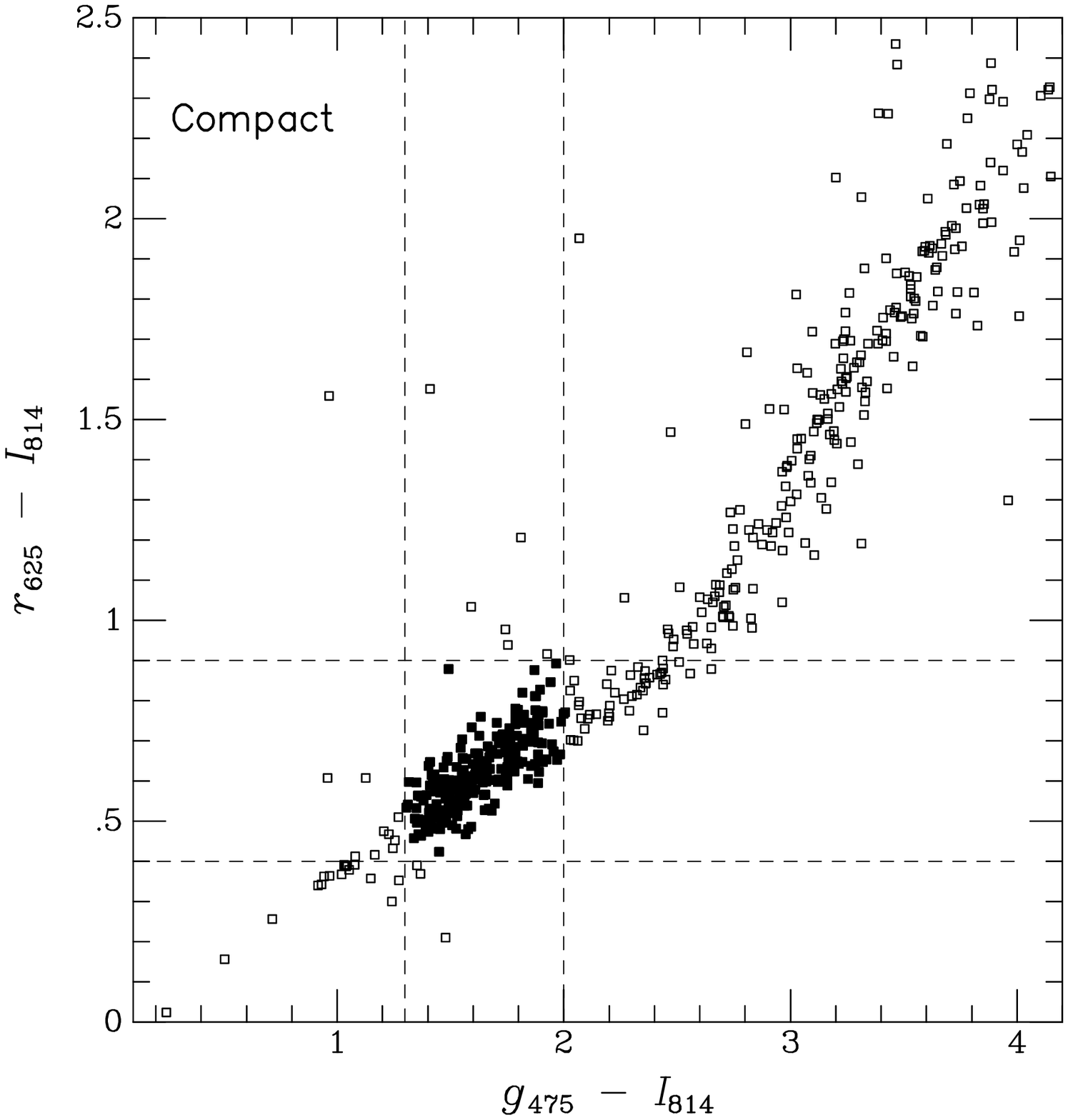}
\caption{Color-color diagram of $\racs-\Iacs$ versus
 $\gacs-\Iacs$ for objects in the \esog\ field with $17<\Iacs<25$ 
(the limit for the 2-D surface photometry fits). 
The left panel shows all objects in this magnitude range, while the right panel shows
``compact'' objects, having the SExtractor parameter
\textsc{class\_star}$\,>\,$0.85.  We do not select based on
 \textsc{class\_star}, but the comparison illustrates the difference
 between ``extended'' and ``compact'' object sequences.  The latter
 includes globular clusters and distant background objects, as well as stars.
The color selection for the UCD candidates is delineated by
the intersection of the horizontal and vertical dashed lines:
$0.4 < \racs-\Iacs < 0.9$ and $ 1.3 < \gacs-\Iacs < 2.0$, and solid
  points are used for objects within this region. }
\label{fig:allcolor}
\end{figure*}

The color-magnitude diagrams in Figure~\ref{fig:all} illustrate our adopted
photometric cuts as applied to the sample of objects detected in the \esog\ field.
Figure~\ref{fig:allcolor} shows the color cuts in the \rIcolor\ versus
\gIcolor\ plane for all objects with $\Iacs<25$.  We plot both
the complete sample of objects (left panel), and the subset with SExtractor
\textsc{class\_star} parameter greater than 0.85 (very compact or stellar
objects; right panel).  Although we do not use \textsc{class\_star} as a selection
criterion, comparison of the plots indicates the location in this diagram of the
likely GCs and UCD candidates.

\begin{figure*}
\epsscale{1.1}
\plottwo{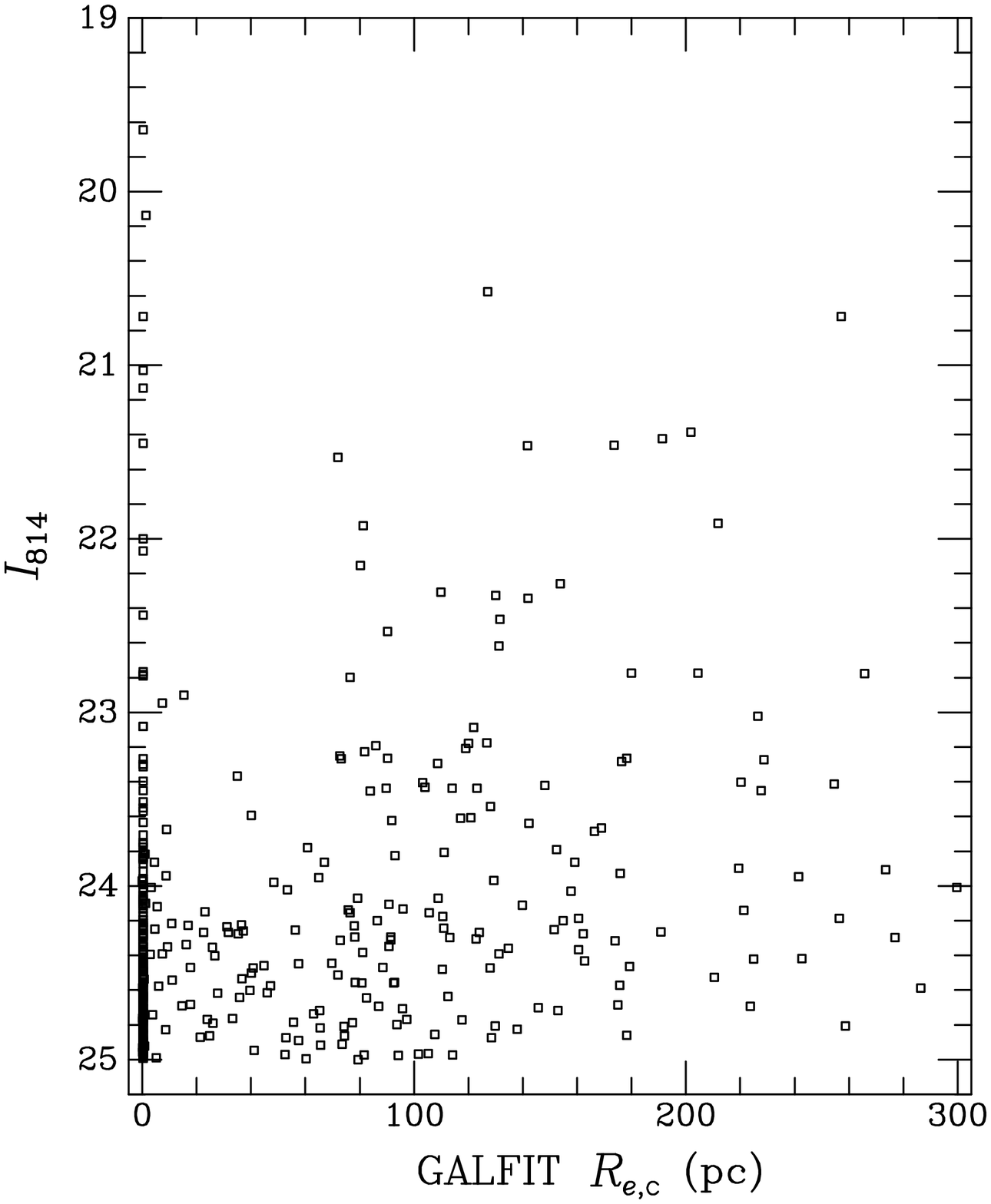}{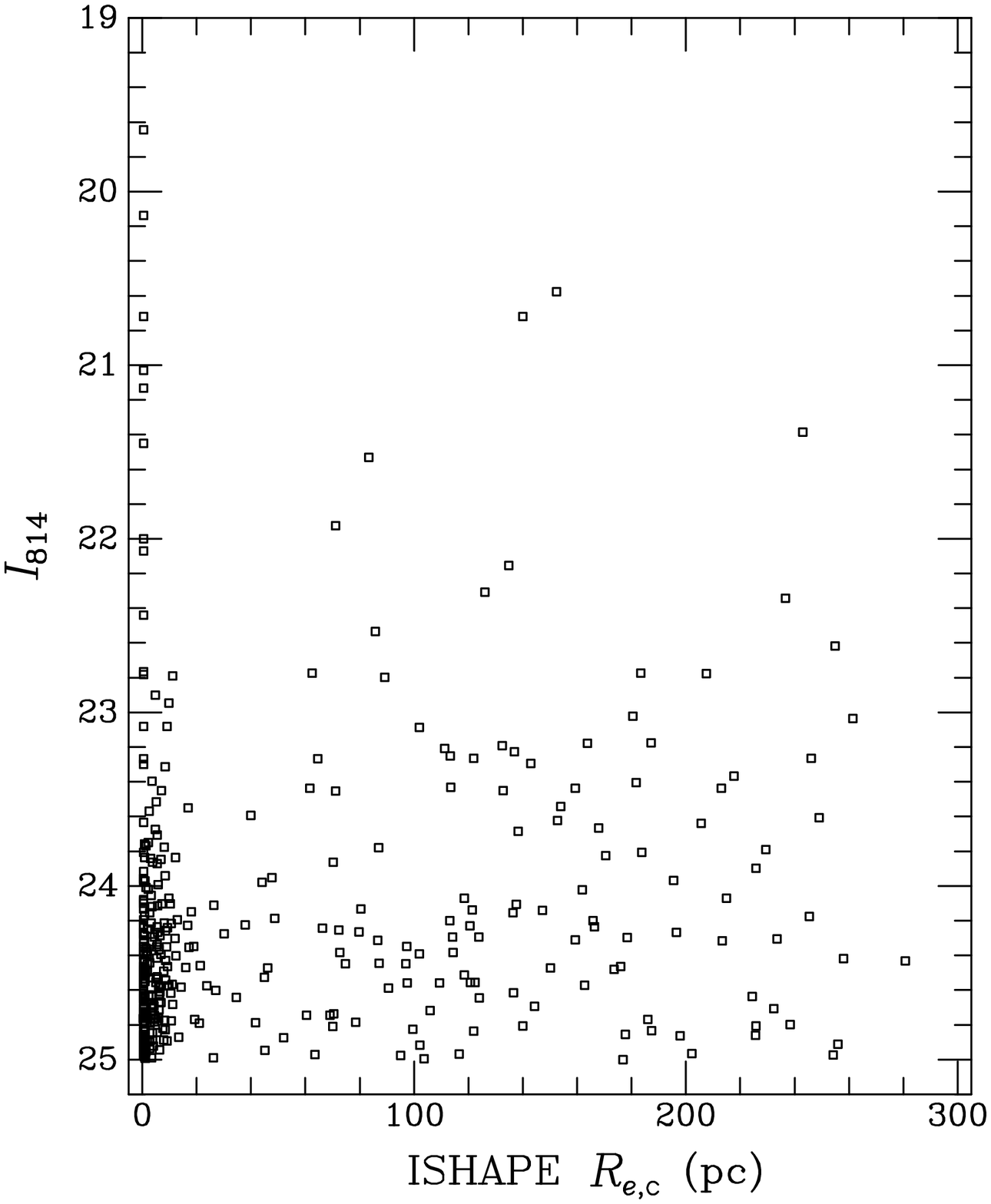}
\caption{Magnitude-size diagrams for the selected sample of objects in the \esog\
  field with $\Iacs<25$ and the color cuts given in the preceding
  figures.  We use SExtractor \magauto\ for \Iacs\ and circular
  half-light radii $\rc$ from the Galfit \sersic\ model (left) and Ishape
  King model (right) fits.  Note that
  $\rc = \rc\sqrt{1{-}\epsilon}\,$, where $\rc$ is the fitted half-light radius along the major
 axis, $\epsilon$ is the fitted ellipticity, and $(1{-}\epsilon)$ the
  axis ratio.
 The image scale at the distance of \esog\ is 33~pc~pix$^{-1\,}$.
}
\label{fig:magsize}
\end{figure*}

\subsection{Size and Shape Measurements}

To measure object sizes, we used the programs Ishape (Larsen 1999) and
Galfit (Peng \etal\ 2002) to model the 2-D profiles of objects in the very
deep F814W image.  Ishape is designed for modeling the light
distributions of marginally resolved sources such as extragalactic GCs, while
Galfit is intended for modeling resolved galaxy light distributions.  It
therefore seemed fitting to use both in a search for UCDs, which straddle the
range between GCs and dwarf galaxies.  For the Ishape fits, we used the
``KING30'' profile, a King (1962) model with concentration parameter $c=30$,
which works well for marginally resolved GCs (e.g., Larsen \& Brodie 2000).
For Galfit, we used a single \sersic\ (1968) model, which has one more degree
of freedom than KING30.  Both programs are quite robust, with typical errors
of 10-15\% for compact but
high signal-to-noise sources such as we have here (see Blakeslee \etal\ 2006;
Barber~DeGraaff \etal\ 2007).  We fitted elliptical models, and use the
circularized effective radius $\rc = R_e\sqrt{q} = R_e\sqrt{1-\epsilon}$,
where $R_e$ is the effective radius along the major axis, $q$ is the fitted
axis ratio, and $\epsilon$ is the ellipticity.

We modeled all objects in the field with $17<\Iacs<25$, within the color
ranges given in \S\,\ref{sec:cmag}, and with SExtractor Kron radius $\leq30$
pix (1~kpc).  The Kron radius selection removes objects much larger than the
UCD and compact elliptical candidates that we are interested in; it should
not exclude any objects in Abell S0740 with \sersic-like profiles and half
light radii below $\sim\,$500~pc (see Graham \& Driver 2005).
Figure~\ref{fig:magsize} shows the magnitude-size diagrams using the \rc\
values from Galfit (left panel) and Ishape (right panel), converted to parsecs
using the adopted distance.  The two panels are similar in overall appearance,
except Ishape resolves many objects with $\rc<10$ pc (0.3~pix) that are not
resolved by Galfit; these are probably mainly globular clusters.  In both
cases, there are about a dozen bright, unresolved objects ($\Iacs<23$, $\rc=0$)
that are most likely stars.

\begin{figure}
\epsscale{1.0}
\plotone{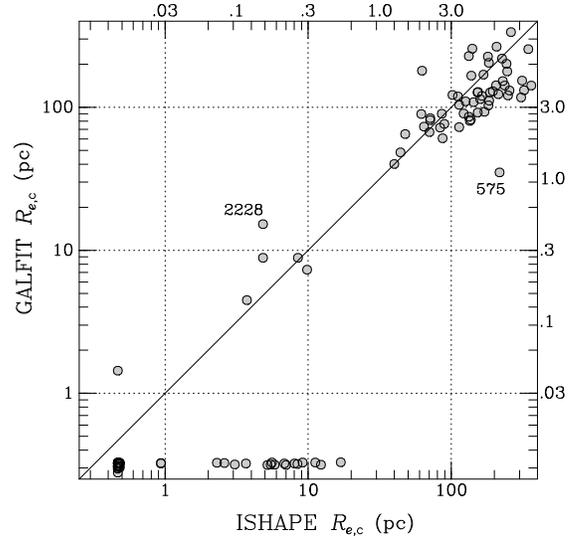}
\caption{Comparison between the circular half-light radii $\rc$ found from the
 Galfit and Ishape fits for objects with
 $\Iacs<24$, the magnitude limit for the UCD selection.  
 Sizes in pixels are plotted along the right and top edges of the figure
 (assuming 33 pc per pixel).
 Ishape is designed for marginally resolved sources and can measure
 sizes for smaller objects, while Galfit can perform more detailed
 analyses of larger objects.
 Two moderate outliers are marked: 2228 is a blend of two objects and
 575 is an edge-on galaxy having a bright subclump; the programs model
 different regions in these two composite sources.
 Otherwise, the two programs agree fairly well, with the exception of objects
 with $\rc\lta10\,$pc (0.3 pix) which Galfit mostly fails to resolve.
\vspace{0.25cm}
}
\label{fig:galshape}
\end{figure}

Figure~\ref{fig:galshape} shows a direct comparison of Galfit and Ishape sizes
for objects with $\Iacs<24$, the magnitude limit for our UCD candidate
selection.  To this limit, the agreement is quite good, apart from the objects
unresolved by Galfit (the agreement worsens for fainter objects).  Ishape does
not do as well for the sizes of larger objects, because it has a limited fit
radius of only a few pixels and overestimates the sizes of larger objects by
about 50\%.  The two worst outliers among the objects that
are resolved by both software packages are irregular objects: \#2228 is a
blended double source, and \#575 is a bright clump within a larger edge-on galaxy.
For the final list of object sizes, we adopted the Galfit \rc\ measurement if
it was greater than 2~pix (66~pc); otherwise, we used the Ishape value for~\rc.

\begin{figure}
\epsscale{1.05}
\plotone{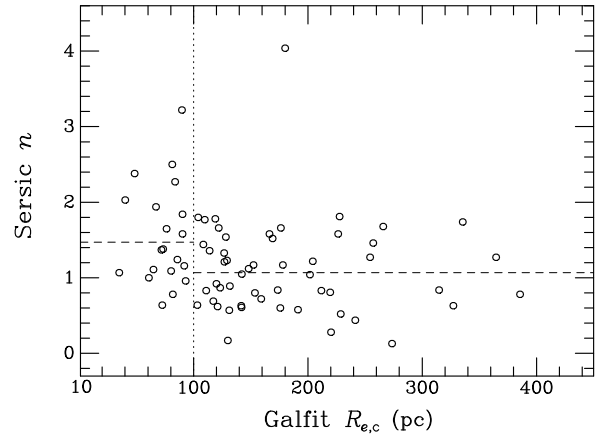}
\caption{\sersic\ index $n$ is plotted against the circularized half light
  radius $\rc$ for the Galfit \sersic\ model fits.  The dashed lines show
  the biweight mean values of $1.47\pm0.15$ and $1.07\pm0.07$ for the
  objects with $10<\rc<100$ pc and $100<\rc<400$ pc, respectively.}
\vspace{0.2cm}
\label{fig:sersic}
\end{figure}

Figure~\ref{fig:sersic} plots the \sersic\ index $n$ against \rc\ from the
Galfit \sersic\ model fits.  Interestingly, the mean $n$ value appears to be
lower for objects with $\rc>100$~pc.  The biweight mean (to reduce the effect
of outliers) is $\langle n \rangle = 1.47\pm0.15$ for objects with \rc =
10-100 pc, and $\langle n \rangle = 1.07\pm0.07$ for $\rc>100$ pc, a
2.4-$\sigma$ difference.  (This includes all objects fitted by Galfit with
sizes in this range, even when the Ishape model was used for the final size.)
The biweight scatters in $n$ for the two groups are 0.66 and 0.51, respectively.
In comparison, the median \sersic\ index for the 21 Virgo and
Fornax UCDs analyzed by Evstigneeva \etal\ (2008) was 2.2, with a large range.
There is a good correspondence between \sersic\ indices measured by Galfit and
morphological type (Blakeslee \etal\ 2006).  Thus, Figure~\ref{fig:sersic} may
indicate that the larger objects in the \esog\ field are dominated by
background galaxies with exponential profiles, while the ones in the 10-100 pc
range include a sizable fraction of UCDs.  Follow-up spectroscopy is necessary
to confirm if this is actually the case.

\begin{figure*}
\epsscale{0.9}
\plotone{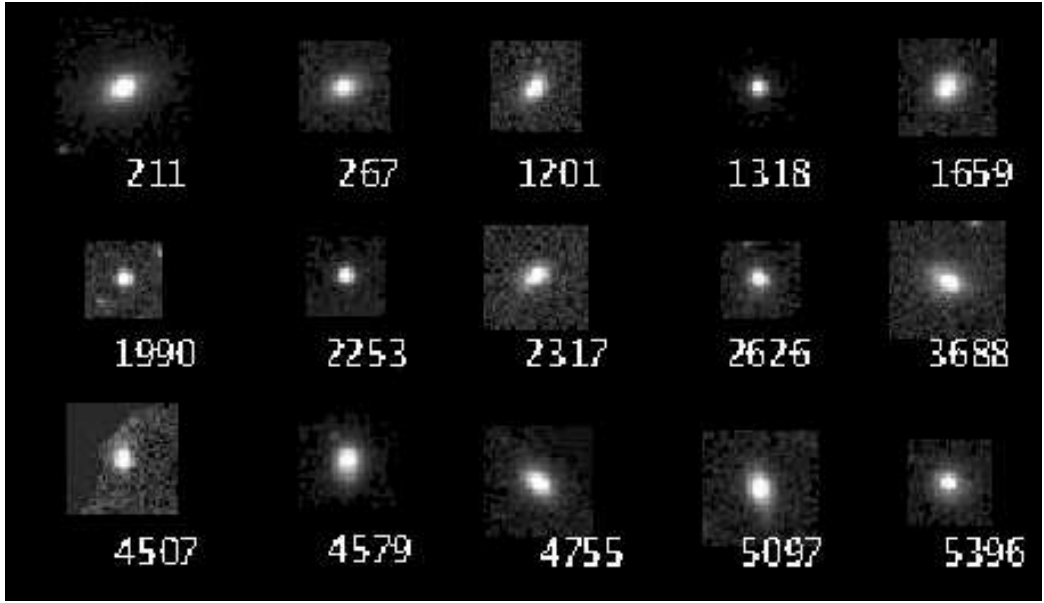}
\caption{%
F814W band images of the candidate ultra-compact dwarf galaxies in the
field of \esog.  These objects meet the color selection criteria, have
$\Iacs<24$, half-light radii in the range 10 to 100~pc, and ellipticity
$\epsilon<0.5$. One other source (575, shown in the following figure) 
ostensibly meeting these criteria was rejected as a subcomponent of 
an elongated edge-on galaxy.  Faint halos of light are visible here
around objects 211, 3688, 4579, and some others; most have such halo light
when examined closely.  Object 4507 is near the edge of a masked
region. 
}
\label{fig:UCDs}
\end{figure*}

\begin{figure*}
\epsscale{0.9}
\plotone{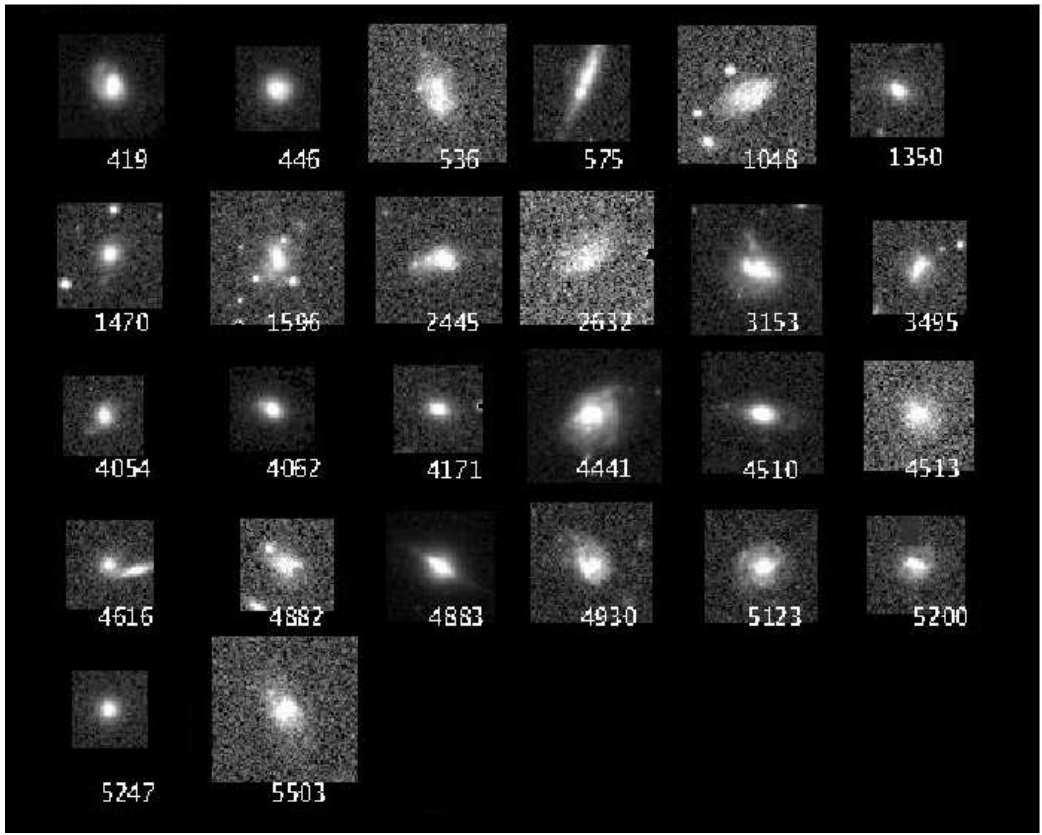}
\caption{F814W band images of objects in the field of \esog\ meeting all
  the selection criteria for UCDs, except having slightly larger sizes
  in the range 100~to 300~pc (plus object 575, noted in the caption to
  Fig.\,\ref{fig:UCDs}). These objects are more irregular in appearance;
  some appear to be background spiral galaxies.}
\label{fig:rest}
\end{figure*}

\begin{figure}
\epsscale{1.03}
\plotone{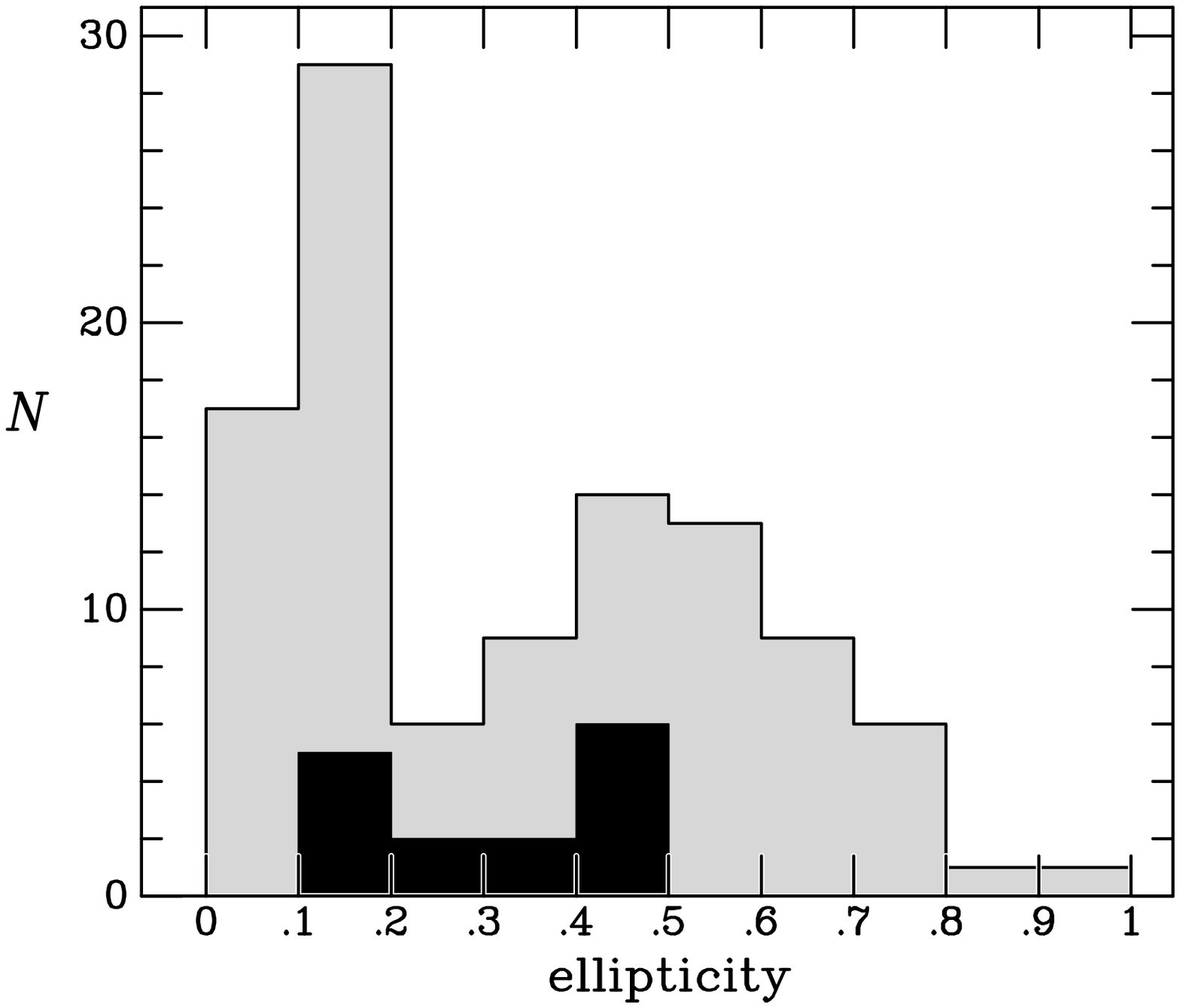}
\caption{Ellipiticity $\epsilon$ distributions for the final sample of 15 UCD
  candidates (black histogram) and all other fitted objects (gray
  histogram) in the same
  magnitude and color ranges ($\Iacs<24$ and color cuts from
  Fig.~\ref{fig:colorage}).  The larger sample is clearly bimodal with
  stellar/GC and extended components.  The UCDs also show two peaks near
  $\epsilon\approx0.17$ and $\epsilon\approx0.45$, although these have only 5 and 6
  objects, respectively, and are subject to small number statistics.
  None of the UCD candidates has an ellipticity $\epsilon<0.16$.
\vspace{0.4cm}
}
\label{fig:ellip}
\end{figure}

\section{Properties of UCD candidates}
\label{sec:props}

For the final sample of most likely UCD candidates, we select all objects with
$\Iacs<24$, 
$1.3 < \gIcolor < 2.0$,
$0.4 < \rIcolor < 0.9$,
10\,pc $<\rc<$ 100\,pc,
and $\epsilon<0.5$ (to eliminate disks and other very elongated objects).
Cut-out images of the 15 UCDs candidates meeting these criteria are displayed in
Figure~\ref{fig:UCDs}.  We removed one object, \#575, from the sample because
it appears to be a subclump of a very elongated galaxy (although it
could be a projection).  All of the
remaining UCD candidates appear to be genuine compact, but nonstellar, early-type
galaxies having colors consistent with being members of Abell~S0740.
When examined closely, many of these objects also show a faint ``fuzz'' of halo
light at radii $r>4$~\rc\ and surface brightness levels 
$\mu_I \approx 24.0$-24.5 mag~arcsec$^{-1}$, which is 
well in excess of PSF blurring effects.

In Figure~\ref{fig:rest}, we show an additional set of 26 compact galaxies meeting
all of the UCD selection criteria except that they have larger sizes in the range
100-300~pc; we also include object \#575 in this figure.   Although some of
these galaxies appear simply to be larger UCD candidates, and we label these as
compact ellipticals (cE), others are irregular galaxies, and a few appear to be
small background spirals.  Table~\ref{tab:dat} lists the positions, magnitudes,
colors, and sizes of the 41 objects in Figures~\ref{fig:UCDs}
and~\ref{fig:rest}.  Magnitudes and colors are corrected for extinction as
described above.  The last column of Table~\ref{tab:dat} reports our classifications for
these objects as UCD (all objects in Fig.~\ref{fig:UCDs}), cE, Sp (spiral), S0
(disky galaxy without obvious spiral structure), Irr (irregular), or clump
(subcomponent of an irregular or interacting system).
The ellipticity
distributions of the 15 UCD candidates and other objects in the field within
the same magnitude and color ranges are shown in Figure~\ref{fig:ellip}.
The UCDs have a mean $\epsilon=0.32$ and a range from 0.16 to 0.46.  The
UCD sample selection excludes objects with $\epsilon>0.5$, but it is
interesting that none has $\epsilon\le0.15$.  This may reflect intrinsic
flattening in the UCDs, since there are many
objects in the larger sample that are found to have lower ellipticity values.

\begin{figure}
\epsscale{1.15}
\plotone{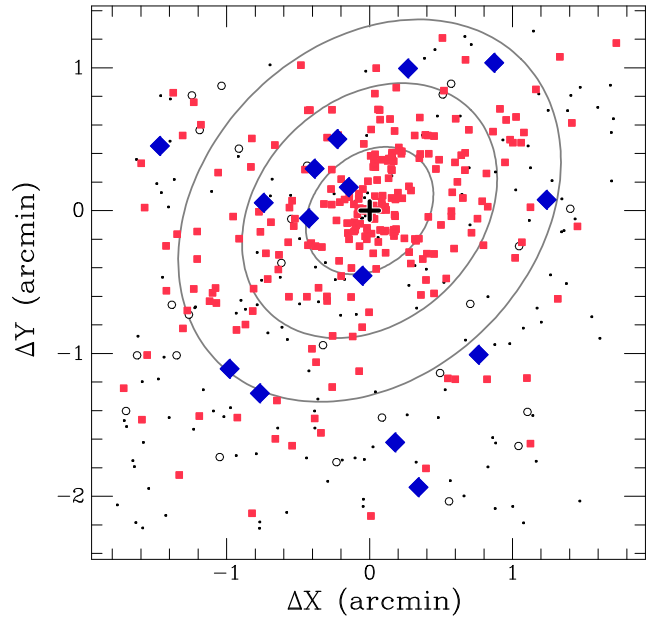}
\caption{Locations of the 15 UCD candidates (blue diamonds), bright 
  globular cluster candidates with $\Iacs<25$ and $\rc<10$\,pc (red
  squares), larger compact galaxies from Fig.\,\ref{fig:rest} (open
  circles), and all other objects in the field with $17<\Iacs<25$ and
  meeting our color cuts (small dots).  The orientation is the same as
  in Fig.\,\ref{fig:pic}, although here we represent the full
  ${\sim\,}3\farcm4{\times}3\farcm4$ field.
  The contours show elliptical isophotes of \esog\ with major axes of
  0\farcm5, 1\farcm0, and 1\farcm5.  The GCs preferentially align along the
  galaxy's major axis.  \hbox{Two-thirds} of the UCD candidates 
  also fall along this direction.
\vspace{0.2cm}
}
\label{fig:positions}
\end{figure}

Figure~\ref{fig:positions} shows the positions of the UCD candidates, larger
compact galaxies, globular clusters candidates with $\Iacs<25$, and other
objects in the field meeting our color and magnitude cuts.  Elliptical
isophotes of \esog\ are also drawn at three radii.
The galaxy is very regular.  It has a mean ellipticity $\epsilon=0.23\pm0.03$
and is oriented $45\degr\pm2\degr$ counter-clockwise from the
$+x$ direction in the observed frame, which translates to a position angle
east of north of
$\hbox{PA} = 66\degr\pm2\degr$. (The errorbars reflect the rms scatter among
the fitted isophotes from the galaxy modeling in Sec.\,\ref{sec:data}.)
A more detailed analysis of the
GC population is in preparation, but we find a best-fit
$\hbox{PA}=71\degr\pm20\degr$ for the GC distribution, in close agreement
with the major axis of the galaxy isophotes.
It is also noteworthy
that 2/3 of the UCD candidates fall along the galaxy's major axis, within
a region covering about 40\% of the image.  Although not statistically
very significant, the UCD alignment along this direction suggests a
link between the UCD and GC populations, and in turn with the stellar
halo of the central elliptical.  We ran a 2-D Kolmogorov-Smirnov test
and found that the spatial distributions of the GCs and UCD candidates
were at least consistent with being the same.  It will be important to
see what fraction of the UCDs lie along the major axis once a
spectroscopically confirmed sample is available.

\begin{figure}
\epsscale{1.05}
\plotone{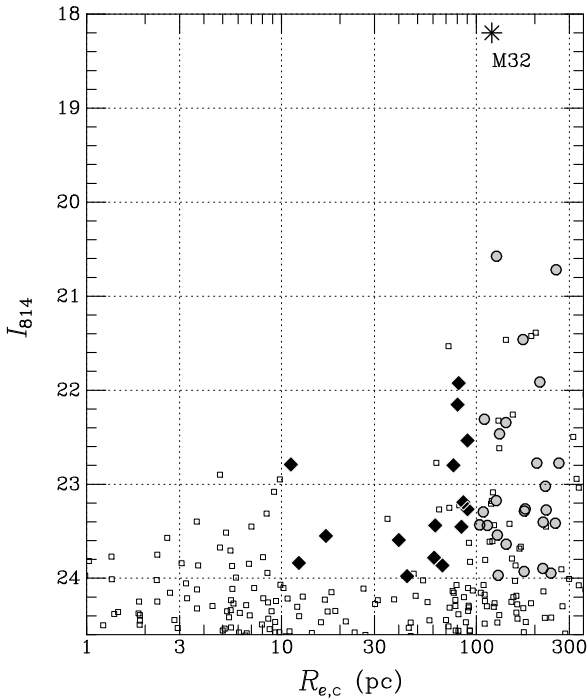}
\caption{F814W magnitude versus size for UCD candidates (filled diamonds),
  larger compact galaxies in the 100-300~pc range (circles) and all
  other objects (open squares) in the \esog\ field that meet our color selection
  criteria and are within the plotted magnitude and size limits.  Objects
  with $\rc{\,<\,}10$~pc are designated globular cluster candidates, while the UCD
  candidates are chosen as having $\rc = 10$ to 100~pc and ellipticity
  $<\,$0.5.  However, there may be a separation between the most compact UCD
  candidates with $\rc<20$~pc, similar to large globular clusters, and those
  with $\rc\gta40$~pc, which may be true compact dwarfs.  Completely
  unresolved objects with $\rc\approx0$ fall off the edge of this logarithmic plot.
  We show the expected location for M32 at this distance; no similar galaxies
  are found in our sample.
\vspace{0.1cm}}
\label{fig:M32}
\end{figure}

In the previous section, we examined the magnitude-size diagrams for the \rc\
values measured separately by Galfit and Ishape.  Figure~\ref{fig:M32}
presents our final magnitude-size diagram using the merged \rc\ values as described
above.  The UCD candidates are marked as large filled diamonds, compact
sources meeting the UCD selection requirements but having \rc\ in the 100-300
pc range are shown as gray circles, and all other objects that were modeled
are shown as open squares.  Taking the UCDs by themselves, or the sample of
UCD and larger compact galaxies together, there is a weak tendency for larger
objects to be brighter.  Formally, our best fit relation for the UCD
candidates implies $\rc\propto L^{0.38\pm0.32}$, but this becomes $\rc\propto
L^{0.53\pm0.25}$ if we omit the most compact candidate with $\rc=11$ pc.
This is consistent with the better determined relation of $\rc\propto
L^{0.68\pm0.13}$ from Evstigeeva \etal\ (2008) using a sample of
confirmed UCDs measured at much higher physical resolution.

There is also an apparent separation in Figure~\ref{fig:M32} between
the three smallest UCDs at $\rc<17$ pc and the other 12
at $\rc>40$ pc.  The first group is very similar to the GCs, while
the latter group appears to blend smoothly with the larger dwarf galaxies.
This may indicate the presence of two distinct types of objects in our UCD
candidate sample, and possibly two different origins for UCDs in general.
However, there is a 17\% probability of a gap as large as the observed one
occurring by chance in this sample.  To our knowledge, no similar gaps have
been reported in previous UCD studies.
Again, spectroscopic confirmation and larger samples of UCDs in diverse
environments are needed to assess the possibility of two distinct populations.
Figure~\ref{fig:M32} also indicates the location that M32 would have in this
diagram if it were at the distance of \esog, using data from Kent (1987).
There are no objects near this position in our sample.
We inspected the images visually to determine if we were somehow missing such
objects in our selection.  One small elliptical located 1\farcm0
approximately due south of \esog\ (at the ``4~o'clock'' position in
Fig.~\ref{fig:pic}) has size and magnitude very close to the expected values
for M32.  However, its colors, $\gIcolor = 2.81$ and $\rIcolor = 0.93$, are
outside our selection range and indicate a higher redshift of $z\approx0.3$.
Thus, we find no M32-like galaxies in this Abell~S0740 field.

\begin{figure}
\epsscale{1.03}
\plotone{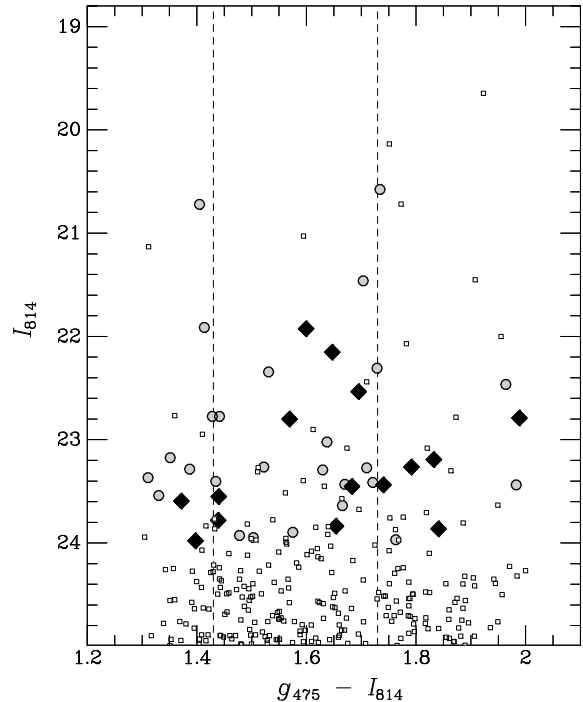}
\caption{Color-magnitude diagram for UCD candidates (diamonds), 
  globular cluster candidates (small squares) and larger compact galaxies
  from Fig.~\ref{fig:rest} (circles).
  The dashed lines indicate the expected locations of the characteristic peaks
  in the globular cluster color distribution.  The UCD candiates are weighted
  toward the red peak location.  It is interesting that most of the brightest
  larger objects (circles at $\Iacs\lta22.8$) lie near the dashed lines.
  The bright objects marked as globular cluster candidates 
  (squares at $\Iacs\lta22.8$) are all unresolved and may be predominantly stars 
  (they all fall off the left edge of Fig.\,\ref{fig:M32}).
\vspace{0.25cm}}
\label{fig:magcolor}
\end{figure}

Figure~\ref{fig:magcolor} presents the final color-magnitude diagram for the
UCD candidates, larger compact galaxies, and GC candidates. The dashed lines
indicate the expected locations of the peaks typically found in GC color
distributions (e.g., Peng \etal\ 2006). 
Past studies have found that UCDs tended to be slightly
redder than the average for the GC population (Wehner \& Harris 2007;
Evstigneeva \etal\ 2008).  Wehner \& Harris described the color-magnitude
sequence for their (unresolved) UCD candidates as an extension of the red GC
subpopulation to higher luminosities.  We also find that the UCDs
in Figure~\ref{fig:magcolor} are weighted towards redder colors, especially
the brightest ones.  We note that several of the compact galaxies with $\rc>100$ pc 
fall intriguingly close to the expected peaks of the GC color distribution,
but the interpretation for these objects is unclear until we have
spectroscopic redshifts.

The $\sim3\farcm4{\,\times\,}3\farcm4$ coverage of our images translates
to about 125${\,\times\,}$125 kpc.
We find that our sample of 15 UCD candidates is reasonable compared with the numbers
found over similar magnitude limits in other clusters. 
In the Fornax cluster, only 4 objects are found
within a similar radius of NGC\,1399 (Mieske \etal\ 2004b; Jones \etal\
2006).  The number in Virgo is complicated by the enormous GC population
around M87, and the lack of a complete high-resolution imaging survey of the
core region.  Jones \etal\ (2006) and Evstigneeva \etal\ (2008) indicate
$\sim\,$5 UCDs within this distance of M87, but an additional five
were found by \hasegan\ \etal\ (2005) in a physical area 70~times
smaller than our own.  Therefore, there may be dozens of UCDs
clustered close to M87.  Perhaps the most comparable survey to ours is
that by Wehner \& Harris (2007) who tabulated 29 UCD candidates in a 5\farcm5
field in the center of the Hydra cluster, to about the same
absolute magnitude limit.  The physical area covered by their imaging
is roughly half of ours; so, we might expect to find $\sim\,$60 candidates
based on this extrapolation.  However, Wehner \& Harris used ground-based
imaging and were not able to select based on objects sizes; if we had not
made the lower \rc\ size cut, we would have a much larger sample of 58
candidates, although the majority of these would be stars and bright GCs.  We
also note that our sample size is roughly comparable to the numbers within
similar radii in the ``Fornax'' and ``Virgo'' numerical models presented by
Bekki \etal~(2003).  

Finally, we provide estimates of the stellar masses of the UCD candidates and
compact galaxies in Figures~\ref{fig:UCDs} and~\ref{fig:rest}.
We calculated the masses of each of the candidates using relations between
mass-to-light ratio $M/L_{V}$ and $\gIcolor$ based on the SSP models from
Figure~\ref{fig:colorage}.  The  $M/L_{V}$ values we find for
the UCD candidates range from $\sim\,$0.5 to 3.5, which are likely uncertain
by about 30-50\%, based on the scatter in the models.  The same uncertainty is
inherent in the stellar mass estimates, which we give in
Table~\ref{tab:dat}.  The values for the UCD candidates range from
$6{\times}10^6$ to $10^8$ \msun, with a median of $3.4{\times}10^7$ \msun.
This agrees well with the dynamically derived masses of
$\sim\,$(2-9)${\,\times\,}10^{7}$ \msun\ from Evstigneeva \etal\ 2007a, a
range that includes 73\% of our candidates.   Similarly, Hilker \etal\ (2007)
found dynamical masses of 1.8 to $9.5{\,\times\,}10^{7}$ \msun\ for five
bright Fornax UCDs.  The two objects with the
largest masses in Table~\ref{tab:dat} (\#419 and \#4883) have \rc\ values too
large for inclusion in the UCD sample, and both have structure indicating 
they are probably background objects (see Fig.~\ref{fig:rest}).  We conclude
that our mass estimates are reasonable for UCDs.

\section{Summary}

We have presented an analysis of three-band ACS/WFC imaging to search for
possible UCDs near the lensing galaxy \esog\ in Abell S0740.  This is an
interesting target for a UCD search because it is a massive central
elliptical in a poor cluster environment with a velocity dispersion similar
to that of Fornax.  We selected objects based on their having magnitudes
brighter than 99\% of the expected GCs population, color
appropriate for an early-type or population II system at this redshift,
ellipticity less than 0.5, and circular half-light radii in the 10-100~pc
range. The radii were measured using both the Galfit and Ishape programs. We
found 15 good UCD candidates meeting the selection criteria, comparable to
the expectations from previous searches.  

In addition, we presented a sample of larger compact galaxies with radii in
the range 100-300~pc, if they are located within the cluster.  These objects
appear to be a mix of irregular background galaxies and larger versions of
the cluster UCDs.  We did not find any counterparts of M32 in this field.
The mean \sersic\ index for the UCD candidates is around 1.5, which is
marginally higher than the value $\sim\,$1 found for the larger compact
galaxies.  This may indicate that the latter objects are dominated by
background disk-like galaxies, while the former group is mainly comprised of
UCDs in the cluster.  Most of the UCD candidates and larger compact galaxies
have visible surrounding halo light, consistent with galaxy threshing models.
There is also evidence that most UCDs are intrinsically flattened, as
none of the 15 UCD candidates has a fitted ellipticity $\epsilon<0.16$.


The magnitude-size and color-magnitude diagrams show general continuity in the
distributions of these parameters from GCs to the UCDs candidates.  
For our limited sample of UCD candidates, we find $\rc\sim L^{0.5}$.  This is
an intriguing proportionality, as it implies a roughly constant surface
brightness for UCDs of different sizes.  The better determined relation from
Evstigneeva \etal\ (2008) is somewhat steeper, but consistent within the errors.
There may be a bifurcation in the UCDs between those with sizes similar
to GCs and larger ones with $\rc>40$~pc, suggesting different origins
for these two groups.  However, because of the small number of objects, the
significance of the observed gap in \rc\ is only 83\%. Therefore, although
suggestive, it remains inconclusive. 
The colors of UCD candidates are weighted towards the red compared to the
expected \gIcolor\ GC color distribution.  Several of the bright compact galaxies with
sizes in the 100-300~pc range have colors near the expected peaks of the
GC color distribution.  It would be useful to know if these objects are
also in the cluster, and what may be their relation to the UCDs.


The majority of UCD candidates align along the major axis of \esog, similar
to the spatial distribution of the bright GCs.  Because of the small numbers involved,
this result is not highly significant, but follow-up spectroscopy
can provide a confirmed sample of UCDs; it will be interesting to see if
these are mainly along the galaxy's major axis.
These findings may appear to support a scenario in which the UCDs are the
high-luminosity extension of the GC system.  However, as discussed in the
Introduction, the true situation is probably more complex, and many red GCs
may actually have their origin as stripped nucleated dwarfs, clouding the
distinction between the main UCD formation scenarios.
It would be useful to discover how the number of UCDs in complete surveys of
many different clusters scales with the GC population of the central galaxy.
We are currently completing a more detailed analysis of the GC population in
this cluster and other similar fields from the same \hst\ program.  We also
plan to obtain spectroscopy for all our UCD candidates to see what fraction of
them are indeed associated with \esog.  The additional information from these
studies should provide further insight into the origin of UCDs and their
connection to the GC populations.

\acknowledgments 
Support for Program number HST-GO-10429 was provided by
NASA through a grant from the Space Telescope Science Institute which is
operated by the Association of Universities for Research in Astronomy,
Incorporated, under NASA contract NAS5-26555.
This research has made use of the NASA/IPAC Extragalactic Database (NED)
which is operated by the Jet Propulsion Laboratory, California Institute
of Technology, under contract with the National Aeronautics and Space
Administration.  RBD wishes to thank Suzanne Hawley and the University of Washington
Department of Astronomy for their hospitality.  We thank Patrick~C\^ot\'e for
helpful discussions, Michael West for incisive comments on the manuscript, and the
anonymous referee for many comments that helped to improve the final
version.  We also thank John Lucey, Russell Smith, and John Tonry, our 
colleagues on the primary \hst\ program.


{}

\vspace{1.2cm}
\clearpage

\LongTables

\begin{deluxetable}{lrrrrrrrrrrrrr}
\tabletypesize{\scriptsize}
\tablewidth{0pt}
\tablecaption{UCD Candidates and Compact Galaxies\label{tab:dat}}
\tablehead{
\colhead{ID} &
\colhead{RA} &
\colhead{Dec} &
\colhead{$I_{814}$} &
\colhead{$\pm$} &
\colhead{$r - I$} &
\colhead{$\pm$} &
\colhead{$g - I$} &
\colhead{$\pm$} &
\colhead{$b / a$\tablenotemark{a}} &
\colhead{$q$\tablenotemark{b}} &
\colhead{$R_{e,{\rm c}}$\tablenotemark{c}} &
\colhead{Mass\tablenotemark{d}} &
\colhead{type\tablenotemark{e}} \\
 &
\colhead{(J2000)} &
\colhead{(J2000)} &
 &
 &
 &
 &
&
 &
 &
 &
\colhead{(pc)} &
\colhead{($M_{\odot}$)} &
}
\startdata
 211 & 205.86130 & -38.18323 &  21.926 & 0.011 &   0.497 &  0.015 &   1.599 &  0.019 &  0.67 &  0.62 &   81.3 & 8.7e+07 & UCD \\ 
 267 & 205.86669 & -38.17410 &  23.438 & 0.014 &   0.631 &  0.029 &   1.740 &  0.043 &  0.75 &  0.68 &   61.6 & 3.4e+07 & UCD \\ 
 419 & 205.86652 & -38.17941 &  21.461 & 0.011 &   0.498 &  0.014 &   1.704 &  0.018 &  0.82 &  0.80 &  173.6 & 1.9e+08 & Sp \\ 
 446 & 205.87906 & -38.15454 &  22.343 & 0.012 &   0.510 &  0.017 &   1.531 &  0.022 &  0.97 &  0.96 &  141.9 & 4.6e+07 & S0 \\ 
 536 & 205.88198 & -38.15171 &  23.273 & 0.019 &   0.734 &  0.048 &   1.710 &  0.056 &  0.66 &  0.57 &  228.7 & 3.6e+07 & Irr \\ 
 575 & 205.86844 & -38.17896 &  23.368 & 0.017 &   0.829 &  0.042 &   1.311 &  0.038 &  0.69 &  0.93 &   35.0 & 8.0e+06 & clump \\ 
1048 & 205.88637 & -38.15403 &  23.403 & 0.020 &   0.829 &  0.048 &   1.434 &  0.045 &  0.57 &  0.56 &  220.3 & 1.2e+07 & Irr \\ 
1201 & 205.88024 & -38.16939 &  23.780 & 0.018 &   0.680 &  0.036 &   1.439 &  0.042 &  0.72 &  0.58 &   60.8 & 8.6e+06 & UCD \\ 
1318 & 205.89067 & -38.15035 &  22.791 & 0.012 &   0.747 &  0.027 &   1.988 &  0.030 &  0.92 &  0.84 &   11.2 & 1.2e+08 & UCD \\ 
1350 & 205.88684 & -38.15907 &  23.433 & 0.017 &   0.825 &  0.038 &   1.670 &  0.049 &  0.71 &  0.56 &  104.0 & 2.8e+07 & cE \\ 
1470 & 205.87982 & -38.17601 &  23.175 & 0.019 &   0.490 &  0.031 &   1.351 &  0.037 &  0.72 &  0.82 &  126.7 & 1.1e+07 & cE \\ 
1596 & 205.88554 & -38.16720 &  23.022 & 0.018 &   0.610 &  0.031 &   1.637 &  0.045 &  0.60 &  0.56 &  226.5 & 3.6e+07 & Irr \\ 
1659 & 205.88557 & -38.16813 &  23.265 & 0.015 &   0.829 &  0.030 &   1.792 &  0.041 &  0.87 &  0.81 &   90.3 & 4.7e+07 & UCD \\ 
1990 & 205.88631 & -38.17262 &  23.837 & 0.017 &   0.567 &  0.033 &   1.654 &  0.042 &  0.96 &  0.81 &   12.3 & 1.8e+07 & UCD \\ 
2253 & 205.87746 & -38.19469 &  23.551 & 0.014 &   0.522 &  0.027 &   1.440 &  0.034 &  0.99 &  0.83 &   16.9 & 1.1e+07 & UCD \\ 
2317 & 205.89302 & -38.16406 &  23.863 & 0.019 &   0.655 &  0.037 &   1.841 &  0.060 &  0.71 &  0.55 &   67.1 & 3.1e+07 & UCD \\ 
2445 & 205.87746 & -38.19764 &  23.264 & 0.018 &   0.579 &  0.032 &   1.522 &  0.045 &  0.53 &  0.58 &  178.2 & 1.9e+07 & Irr \\ 
2626 & 205.89274 & -38.16961 &  23.594 & 0.016 &   0.660 &  0.031 &   1.372 &  0.034 &  0.88 &  0.70 &   40.0 & 7.9e+06 & UCD \\ 
2632 & 205.89377 & -38.16771 &  23.947 & 0.033 &   0.876 &  0.087 &   1.503 &  0.093 &  0.51 &  0.56 &  241.5 & 9.5e+06 & Irr \\ 
3153 & 205.88536 & -38.19365 &  21.912 & 0.012 &   0.499 &  0.016 &   1.413 &  0.020 &  0.87 &  0.66 &  211.8 & 4.4e+07 & Irr \\ 
3495 & 205.90039 & -38.16848 &  23.638 & 0.021 &   0.692 &  0.043 &   1.665 &  0.059 &  0.73 &  0.59 &  142.2 & 2.3e+07 & Irr \\ 
3688 & 205.89787 & -38.17786 &  22.535 & 0.015 &   0.640 &  0.022 &   1.695 &  0.034 &  0.77 &  0.54 &   90.2 & 6.9e+07 & UCD \\ 
4054 & 205.89599 & -38.19076 &  23.541 & 0.018 &   0.549 &  0.033 &   1.330 &  0.041 &  0.75 &  0.80 &  128.1 & 7.2e+06 & Sp \\ 
4062 & 205.91205 & -38.15832 &  22.308 & 0.011 &   0.554 &  0.022 &   1.729 &  0.024 &  0.76 &  0.65 &  109.9 & 9.4e+07 & cE \\ 
4171 & 205.91251 & -38.16059 &  23.294 & 0.015 &   0.513 &  0.037 &   1.630 &  0.042 &  0.73 &  0.62 &  108.6 & 2.7e+07 & cE \\ 
4441 & 205.90959 & -38.17646 &  20.721 & 0.010 &   0.531 &  0.012 &   1.405 &  0.013 &  0.78 &  0.76 &  257.2 & 1.3e+08 & Sp/Irr \\ 
4507 & 205.90258 & -38.19380 &  23.979 & 0.020 &   0.664 &  0.044 &   1.397 &  0.052 &  0.68 &  0.54 &   44.1 & 6.1e+06 & UCD \\ 
4510 & 205.91878 & -38.16098 &  22.464 & 0.012 &   0.795 &  0.029 &   1.964 &  0.035 &  0.50 &  0.56 &  131.6 & 1.5e+08 & cE \\ 
4513 & 205.92092 & -38.15667 &  23.897 & 0.028 &   0.543 &  0.075 &   1.575 &  0.083 &  0.85 &  0.89 &  219.5 & 1.3e+07 & Irr \\ 
4579 & 205.91782 & -38.16729 &  22.154 & 0.011 &   0.767 &  0.017 &   1.647 &  0.020 &  0.76 &  0.77 &   80.2 & 8.3e+07 & UCD \\ 
4616 & 205.90719 & -38.19038 &  23.969 & 0.026 &   0.625 &  0.055 &   1.763 &  0.084 &  0.85 &  0.90 &  129.3 & 2.3e+07 & clump \\ 
4755 & 205.91962 & -38.17162 &  22.799 & 0.012 &   0.534 &  0.020 &   1.569 &  0.028 &  0.73 &  0.54 &   76.4 & 3.5e+07 & UCD \\ 
4882 & 205.92918 & -38.15779 &  23.928 & 0.025 &   0.509 &  0.065 &   1.478 &  0.069 &  0.79 &  0.76 &  175.8 & 8.8e+06 & Irr \\ 
4883 & 205.90786 & -38.20153 &  20.577 & 0.010 &   0.709 &  0.012 &   1.734 &  0.013 &  0.69 &  0.54 &  127.1 & 4.7e+08 & S0 \\ 
4930 & 205.91646 & -38.18593 &  22.775 & 0.015 &   0.512 &  0.024 &   1.441 &  0.031 &  0.73 &  0.82 &  204.4 & 2.2e+07 & Irr \\ 
5097 & 205.91919 & -38.18841 &  23.193 & 0.014 &   0.860 &  0.030 &   1.832 &  0.042 &  0.64 &  0.59 &   85.9 & 5.6e+07 & UCD \\ 
5123 & 205.91309 & -38.20199 &  22.776 & 0.015 &   0.773 &  0.029 &   1.428 &  0.033 &  0.90 &  0.90 &  265.8 & 2.1e+07 & Sp \\ 
5200 & 205.93059 & -38.16994 &  23.284 & 0.017 &   0.863 &  0.054 &   1.387 &  0.040 &  0.84 &  0.62 &  176.4 & 1.1e+07 & Irr \\ 
5247 & 205.92507 & -38.18285 &  23.438 & 0.017 &   0.693 &  0.045 &   1.983 &  0.061 &  0.93 &  0.98 &  114.0 & 6.5e+07 & cE \\ 
5396 & 205.92415 & -38.19283 &  23.454 & 0.014 &   0.544 &  0.036 &   1.683 &  0.042 &  0.92 &  0.83 &   83.9 & 2.8e+07 & UCD \\ 
5503 & 205.92445 & -38.19675 &  23.414 & 0.023 &   0.489 &  0.056 &   1.721 &  0.071 &  0.66 &  0.72 &  254.5 & 3.3e+07 & Irr \\
\enddata
%
\tablenotetext{a}{Axis ratio measured by SExtractor; no PSF correction.}
\tablenotetext{b}{Intrinsic axis ratio $q=1-\epsilon$ from our 2-D modeling with PSF correction.}
\tablenotetext{c}{Fitted circularized effective radius $\rc=R_e\sqrt{q}.$}
\tablenotetext{d}{Photometrically derived stellar mass estimate.}
\tablenotetext{e}{\vspace{-1cm}Morphological type from our visual inspection.
  All objects in Fig.~\ref{fig:UCDs} are type UCD; see text for further details.}
\end{deluxetable}

\end{document}